\documentclass[aps,preprint,showpacs,preprintnumbers,amsmath,amssymb]{revtex4}
\usepackage{amsmath,mathrsfs,amsbsy,color,graphicx,bm,amsthm,amsfonts}
\usepackage{units}
\usepackage{bbm}
\usepackage{times}
\usepackage{dcolumn}
\usepackage{mathrsfs}
\usepackage{amsmath,amssymb,epsfig}
\usepackage{amsmath}
\newcommand{\udots}{\mathinner{\mskip1mu\raise1pt\vbox{\kern7pt\hbox{.}}
\mskip2mu\raise4pt\hbox{.}\mskip2mu\raise7pt\hbox{.}\mskip1mu}}
\begin{document}
\title{ Advantages and disadvantages of maximally entangled states in dilaton black hole background }
\author{Zhen Yang \footnote{Email: yangzhen@lnnu.edu.cn}, He Cheng,  Si-Han Li \footnote{Email: sihanli123456@163.com (corresponding author)}}
\affiliation{$^1$ Department of Physics, Liaoning Normal University, Dalian 116029, China}


\begin{abstract}
We investigate quantum entanglement and coherence for four classes of Bell-like fermionic states in the vicinity of the event horizon of a  Garfinkle-Horowitz-Strominger (GHS) dilaton black hole. Contrary to the common expectation that maximally entangled states always provide superior quantum resources, our results show that their entanglement can be lower than that of suitably chosen non-maximally entangled states in this curved spacetime background. This reveals that non-maximally entangled states may offer operational advantages for entanglement-based tasks under gravitational effects.
In contrast, quantum coherence exhibits monotonic behavior: larger initial coherence leads to systematically enhanced robustness against the dilaton induced degradation.  These results indicate that the optimal choice of initial quantum states depends sensitively on the specific quantum resource, either quantum entanglement or quantum coherence, required for quantum information processing near a dilaton black hole.
\end{abstract}

\vspace*{0.5cm}
\maketitle
\section{Introduction}
Quantum coherence, rooted in the superposition principle, constitutes one of the most elementary signatures distinguishing quantum systems from their classical counterparts \cite{L1}. Its creation and maintenance are critical for implementing quantum information processing tasks \cite{L2}. While coherence has long been recognized as central to quantum theory, it gained formal quantification only after Baumgratz $et$ $al$. introduced a resource-theoretic framework, defining measures such as the \( l_1 \)-norm  and relative entropy of coherence \cite{L47}. Analogous to entanglement, coherence constitutes a valuable quantum resource, playing significant roles in areas including quantum communication networks, teleportation, quantum batteries, quantum computation, and quantum biology \cite{L4,L5,L6,L7,L8,L9,L10,L11,L12,L13,L13-1}. It is a property not limited to multipartite systems, appearing from single-qubit superpositions to complex multipartite correlations \cite{L2,N1}. Importantly, entanglement, which arises from nonlocal superpositions, can be generated from quantum coherence without introducing additional resources. This relationship implies that an entanglement monotone can define a corresponding coherence monotone for a given state \cite{L14,W1}. Despite substantial progress in understanding the interplay between coherence and entanglement, it continues to be an active subject of research.

Quantum information in gravitational settings, which unifies concepts from quantum information theory, quantum field theory, and general relativity, has attracted considerable attention in recent years \cite{SDF1,SDF2,SDF3,SDF4,SDF5,SDF6,SDF8,SDF9,SDF10,SDF12,SDF13,SDF16,SDF18,SDF19,SDF21,SDF22,SDF23,SDF24,SDF26,SDF27,SDF28,SDF29,SDF30,SDF31,SDF32,SDF33,SDF35,SDF36,SDF37,SDF38,SDF39,SDF40,SDF41,SDF42,SDF43,SDF44,SDF46,SDF47,SDF48,SDF49,SDF50,SDF51,SDF52,QYM1,QYM2,SDF53,SDF54,SDF55,SDF56,SDF57,SDF58,SDF59,SDF60,SDF61,SDF62,SDF63,SDF64,SDF65,QTP1,QTP2,QTP3,QTP4,QTP5,QTP6,QTP7,QTP8,QTP9,QTP10,wrc1}. Research in this area generally focuses on two complementary directions: (i) employing quantum technologies to probe the fundamental structure of spacetime, and (ii) investigating how gravitational effects impact quantum resources such as entanglement and coherence. Within this context, string theory predicts that dilaton fields modify the characteristics of black holes \cite{J9,J10,J11,J12}. Notably, the GHS dilaton black hole exhibits a Hawking radiation spectrum that depends not only on the black hole mass but also on the dilaton field, which itself contributes to the gravitational dynamics. Previous studies have explored the influence of the dilaton black hole background on various quantum phenomena, including quantum correlations, coherence, and entropic uncertainty relations \cite{J44,J45,J46,J47,J48,J49,J50,J51,J52,J53,J54,J55,J56,J57,J58,J59,J60}. These works generally indicate that maximally entangled states provide stronger quantum resources than non-maximally entangled states. However, the experimental preparation of maximally entangled states remains challenging, making non-maximally entangled states more accessible for practical quantum information protocols in curved spacetime. This observation naturally leads to a fundamental question: under the influence of the dilaton black hole, could non-maximally entangled states actually outperform maximally entangled states in terms of resource efficiency?
Motivated by this question, we investigate the behavior of four classes of Bell-like fermionic states near the event horizon of a GHS black hole. Our aim is twofold: first, to compare the robustness of entanglement and coherence across different initial states under gravitational effects; second, to identify the Bell-like state that retains the largest residual entanglement and coherence, providing a potential candidate for optimized quantum information processing in a curved spacetime background. This study not only addresses practical considerations in quantum state preparation but also deepens our understanding of how gravitational fields modulate quantum resources.

In this work, we investigate the influence of Hawking radiation on quantum entanglement and coherence for four classes of Bell-like fermionic states in the spacetime of a GHS dilaton black hole. We begin by assuming that Alice and Bob initially share these Bell-like states in flat Minkowski spacetime. Both observers are then placed near the event horizon of the black hole, allowing us to examine the effects of spacetime curvature on their shared quantum resources. We analytically derive expressions for fermionic entanglement and coherence for each class of Bell-like states in the curved background and analyze how the Hawking effect modifies these quantities. Our results reveal several noteworthy features. First, the entanglement of certain non-maximally entangled states can surpass that of maximally entangled states under the gravitational influence of the dilaton black hole. Second, in contrast to entanglement, larger initial quantum coherence consistently provides greater robustness, maintaining higher resource values in curved spacetime. Third, the degradation patterns of entanglement differ among the Bell-like states: for two classes, entanglement may undergo sudden death, whereas for the remaining two, it persists indefinitely despite the Hawking radiation. These findings highlight the importance of selecting appropriate initial states according to the type of quantum resource required. In particular, optimizing either entanglement or coherence in the presence of gravitational effects demands careful consideration of both the initial state and the underlying spacetime geometry, offering valuable guidance for quantum information processing in curved spacetime scenarios.

The paper is organized as follows. In Sec.~II, we present the quantization of Dirac fields in the spacetime of a GHS dilaton black hole. In Sec.~III, we analyze fermionic entanglement and coherence for four classes of Bell-like states in this background. Finally, Sec.~VI summarizes our main results and conclusions.

\section{Quantization of Dirac field in dilaton black hole }
In this section, we reformulate the quantization of a massless Dirac field in the spacetime of a GHS dilaton black hole. Our aim is to construct the corresponding mode functions both outside and inside the event horizon and to establish the Bogoliubov transformations between the Kruskal and dilaton mode operators, which encode the Hawking effect for fermionic fields in this background.  The evolution of a massless  field on a generic curved background is governed by the covariant Dirac equation \cite{J61,J62}
\begin{eqnarray}\label{R13}
[\gamma^a e^\mu_a(\partial_\mu + \Gamma_\mu)]\Psi = 0,
\end{eqnarray}
where $\gamma^a$ are the flat-space Dirac matrices,  $e^\mu_a$ is the inverse tetrad, and
$\Gamma_\mu = \frac{1}{8}[\gamma^a,\gamma ^b]e^\nu_a e_{b\nu;\mu}$ denotes the spin connection.
The tetrad is defined through the standard decomposition of the metric, $e^a_\mu$ is defined by $g_{\mu\nu} = \eta_{ab} e^a_\mu e^b_\nu$ with $\eta_{ab} = \rm{diag}(-1,1,1,1)$.

The geometry of the GHS dilaton black hole is described by the static, spherically symmetric metric
\begin{eqnarray}\label{R14}
ds^2 = -\bigg(\frac{r-2M}{r-2D}\bigg)dt^2 + \bigg(\frac{r-2M}{r-2D}\bigg)^{-1} dr^2 + r(r-2D)d\Omega^2,
\end{eqnarray}
where $M$ and $D$ represent the mass and the dilaton of the black hole, respectively \cite{J44}. Throughout this work we adopt natural units $\hbar = G = c = \kappa_B = 1$. The corresponding Hawking temperature is controlled by the surface gravity at the outer horizon and takes the fermionic Fermi-Dirac form  $T = \frac{1}{8\pi (M -D)}$, which governs the thermal excitation spectrum of outgoing spinor modes near the horizon
\cite{J49,J50,J51}.
To prepare for the separation of variables in the Dirac equation, we introduce the following diagonal tetrad field,
\begin{eqnarray}\label{R15}
e^a_\mu = \rm{diag}\bigg(\sqrt{f},\frac{1}{\sqrt{f}},\sqrt{r\tilde{r}},\sqrt{r\tilde{r}} \sin \theta\bigg),
\end{eqnarray}
where $f=\frac{(r-2M)}{\tilde{r}}$ and $\tilde{r}=r-2D$. Inserting this tetrad into the massless Dirac equation yields the explicit curved-spacetime equation
\begin{eqnarray}\label{R16}
&&-\frac{r_0}{\sqrt{f}} \frac{\partial \Psi}{\partial t} + \sqrt{f} \gamma_1\bigg(\frac{\partial}{\partial r} + \frac{r-D}{r \tilde{r}} + \frac{1}{4f} \frac{df}{dr}\bigg)\Psi\nonumber\\
 &&+\frac{\gamma_2}{\sqrt{r\tilde{r}}}\bigg(\frac{\partial}{\partial \theta}+\frac{\cot \theta}{2}\bigg)\Psi + \frac{\gamma_3}{\sqrt{r\tilde{r}}\sin \theta} \frac{\partial \Psi}{\partial \varphi} = 0.
\end{eqnarray}
If we use $\Psi = f^{-\frac{1}{4}} \Phi$, we can solve the Dirac equation near the event horizon of the black hole \cite{J46}.
For  the exterior region and  interior region of the event horizon, the positive-frequency outgoing solutions behave asymptotically as
\begin{eqnarray}\label{R17}
\Psi^+_{out,\boldmath{k}} = \mathcal{J} e^{-i \omega \mathcal{O}},
\end{eqnarray}
\begin{eqnarray}\label{R18}
\Psi^+_{in,\boldmath{k}} = \mathcal{J} e^{i \omega \mathcal{O}},
\end{eqnarray}
where $\mathcal{O} = t - r_*$, $\mathcal{J}$ is a four-component Dirac spinor, and $\boldmath{k}$ is the wave vector that can be used to label the modes.
Using the above mode functions, the Dirac field can be expanded in the dilaton-mode basis as
\begin{eqnarray}\label{R19}
\Psi = \sum_{\sigma} \int d \boldmath{k}[\hat{a}^{\sigma}_{\boldmath{k}} \Psi^+_{\sigma, \boldmath{k}} + \hat{b}^{\sigma \dagger}_{\boldmath{k}} \Psi^-_{\sigma, \boldmath{k}}],
\end{eqnarray}
where $\sigma = (in, out)$ distinguishes modes supported inside and outside the event horizon. The operators obey the standard canonical anticommutation relations $\hat{a}^{\sigma}_{\boldmath{k}}$ and $\hat{b}^{\sigma \dagger}_{\boldmath{k}}$ 
$\left\{\hat{a}^{out}_{\boldmath{k}},\hat{a}^{out \dagger}_{\boldmath{k'}}\right\} = \left\{\hat{a}^{in}_{\boldmath{k}},\hat{a}^{in \dagger }_{\boldmath{k'}}\right\} = \left\{\hat{b}^{out}_{\boldmath{k}},\hat{b}^{out \dagger}_{\boldmath{k'}}\right\} = \left\{\hat{b}^{in}_{\boldmath{k}},\hat{b}^{in \dagger }_{\boldmath{k'}}\right\} = \delta_{\boldmath{k} \boldmath{k'}}$. 
The corresponding dilaton vacuum $\vert 0 \rangle_D$ is annihilated by all $\hat{a}^{\sigma}_{\boldmath{k}}$.

The Kruskal extension of the GHS dilaton black hole provides a globally regular coordinate system in which the field modes are analytic across the horizon. In this framework, one constructs a complete orthonormal basis of positive-frequency spinor solutions-commonly referred to as the Kruskal modes, originally introduced by Damour and Ruffini \cite{J63}. These modes are particularly suitable for describing the vacuum state perceived by freely falling observers. To express the Dirac field in this regular basis, we expand it as
\begin{eqnarray}\label{R20}
\Psi = \sum_{\sigma} \int d \boldmath{k} \frac{1}{\sqrt{2 \cosh (4 \pi (M - D)\omega)}} \left[ \hat{c}^{\sigma}_{\boldmath{k}} \Phi^+_{\sigma, \boldmath{k}} + \hat{d}^{\sigma \dagger}_{\boldmath{k}} \Phi^-_{\sigma, \boldmath{k}}\right],
\end{eqnarray}
where $\hat{c}^{\sigma}_{\boldmath{k}}$ and $\hat{d}^{\sigma \dagger}_{\boldmath{k}}$ are the fermion annihilation and antifermion creation operators acting on the Kruskal vacuum, respectively \cite{J58,J60}.
Comparing the above expansion with the dilaton-mode decomposition in Eq.(\ref{R19}), one observes that the same quantum field has been expressed in two inequivalent mode bases:
(i) the dilaton modes, naturally associated with static observers outside the horizon, and
(ii) the Kruskal modes, well behaved across the entire spacetime.
This mismatch in the notion of positive frequency gives rise to the Bogoliubov transformations connecting the two bases. Explicitly, the fermionic Bogoliubov transformations relating the Kruskal operators  and the dilaton operators  take the forms
\begin{eqnarray}\label{R21}
\hat{c}^{out}_{\boldmath{k}} = \frac{1}{\sqrt{e^{- 8 \pi (M - D) \omega } + 1} } \hat{a}^{out}_{\boldmath{k}} - \frac{1}{\sqrt{e^{8 \pi (M - D) \omega } + 1} } \hat{b}^{out \dagger}_{\boldmath{k}},\notag\\
\hat{c}^{out \dagger}_{\boldmath{k}} = \frac{1}{\sqrt{e^{- 8 \pi (M - D) \omega } + 1} } \hat{a}^{out \dagger}_{\boldmath{k}} - \frac{1}{\sqrt{e^{8 \pi (M - D) \omega } + 1} } \hat{b}^{out}_{\boldmath{k}}.
\end{eqnarray}
Because the GHS spacetime naturally separates into an exterior (accessible) region and an interior (inaccessible) region, the Kruskal vacuum is perceived by an exterior observer as an entangled state between these two sectors \cite{J59}.
Performing the standard fermionic normalization, one obtains the Kruskal vacuum and single-particle states expressed in the dilaton-mode Fock basis
\begin{eqnarray}\label{R22}
\nonumber&\vert 0 \rangle_K = \frac{1}{\sqrt{e^{- 8 \pi (M - D) \omega } + 1} } \vert 0 \rangle_{out} \vert 0 \rangle_{in} + \frac{1}{\sqrt{e^{8 \pi (M - D) \omega } + 1} } \vert 1 \rangle_{out} \vert 1 \rangle_{in},\\
&\vert 1 \rangle_K = \vert 1 \rangle_{out} \vert 0 \rangle_{in},
\end{eqnarray}
where $\left\{\vert n \rangle_{out}\right\} $ and $\left\{\vert n \rangle_{in}\right\} $
denote fermionic and  antifermionic  number states associated with exterior and interior regions of the horizon, respectively \cite{J45}.

\section{Quantum entanglement and coherence for different types of Bell-like states in dilaton black hole}
\begin{figure}
\centering
\includegraphics[height=2in,width=3.5in]{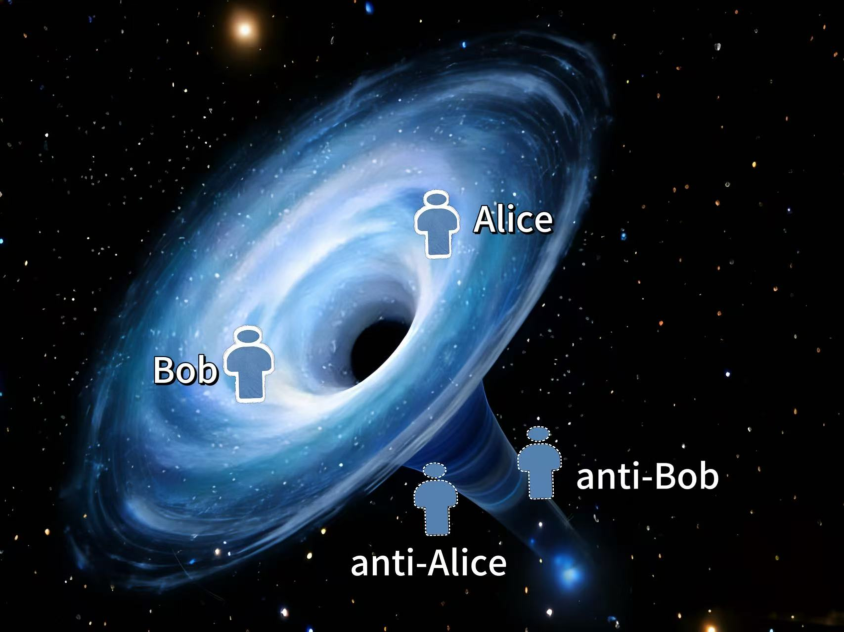}
\caption{Embedding diagram of the black hole and the detector configuration. Alice and Bob hover at fixed radial positions outside the event horizon, while the anti-Alice and anti-Bob modes correspond to the causally inaccessible region inside the  event horizon.}
\label{F0}
\end{figure}

The four different types of Bell-like states of the entangled fermionic modes in the asymptotically flat region of a dilaton black hole can be described as
\begin{eqnarray}\label{w23}
|\phi^{{1},{\pm}}_{AB}\rangle=\alpha|0_{A}\rangle|0_{B}\rangle \pm \sqrt{1-\alpha^{2}}|1_{A}\rangle|1_{B}\rangle,
\end{eqnarray}
\begin{eqnarray}\label{w24}
|\Psi^{{2},{\pm}}_{AB}\rangle=\alpha|0_{A}\rangle|1_{B}\rangle \pm \sqrt{1-\alpha^{2}}|1_{A}\rangle|0_{B}\rangle,
\end{eqnarray}
where the subscripts $A$ and $B$ denote the modes associated with Alice and Bob, respectively.
To accurately characterize the modification of the shared quantum state induced by the Hawking effect, it is necessary to specify both the causal structure of the black hole spacetime and the operational
placement of the detectors. As illustrated in Fig.\ref{F0}, the black hole geometry is divided by the
event horizon into two causally disconnected regions: an exterior region that extends smoothly to an
asymptotically flat infinity, and an interior region from which no information can propagate to outside
observers. The event horizon thus acts as a one-way causal boundary. In our configuration, both Alice
and Bob hover just outside the event horizon and have access only to exterior field modes. The modes
residing inside the horizon are denoted by anti-Alice and anti-Bob, representing degrees of freedom
that are causally disconnected from the exterior region and therefore inaccessible to Alice and Bob.
Consequently, these interior modes must be traced out when constructing the reduced quantum state
relevant for physical observations outside the horizon. Employing Eq.(\ref{R22}), we can rewrite Eqs.(\ref{w23}) and (\ref{w24}) using dilaton modes for both Alice and Bob as
\begin{eqnarray}\label{w25}
|\phi^{{1},{\pm}}_{A{\bar A}B{\bar B}}\rangle&=&
\alpha \bigg(
\frac{1}{\sqrt{e^{- 8 \pi (M - D) \omega_{A} } + 1} } \frac{1}{\sqrt{e^{- 8 \pi (M - D) \omega_{B} } + 1} }|0\rangle_{A}|0\rangle_{\bar A}|0\rangle_{B}|0\rangle_{\bar B}\nonumber\\&&
+\frac{1}{\sqrt{e^{- 8 \pi (M - D) \omega_{A} } + 1} } \frac{1}{\sqrt{e^{ 8 \pi (M - D) \omega_{B} } + 1} }|0\rangle_{A}|0\rangle_{\bar A}|1\rangle_{B}|1\rangle_{\bar B}\nonumber\\&&
+\frac{1}{\sqrt{e^{ 8 \pi (M - D) \omega_{A} } + 1} } \frac{1}{\sqrt{e^{- 8 \pi (M - D) \omega_{B} } + 1} }|1\rangle_{A}|1\rangle_{\bar A}|0\rangle_{B}|0\rangle_{\bar B}\nonumber\\&&
+\frac{1}{\sqrt{e^{ 8 \pi (M - D) \omega_{A} } + 1} } \frac{1}{\sqrt{e^{ 8 \pi (M - D) \omega_{B} } + 1} }|1\rangle_{A}|1\rangle_{\bar A}|1\rangle_{B}|1\rangle_{\bar B} \bigg)\nonumber\\&&\pm
\sqrt{1-\alpha^{2}}|1\rangle_{A}|0\rangle_{\bar A}|1\rangle_{B}|0\rangle_{\bar B},
\end{eqnarray}
\begin{eqnarray}\label{w26}
|\Psi^{{2},{\pm}}_{A{\bar A}B{\bar B}}\rangle&=&
  \frac{\alpha}{\sqrt{e^{- 8 \pi (M - D) \omega_{A} } + 1} }|0\rangle_{A}|0\rangle_{\bar A}|1\rangle_{B}|0\rangle_{\bar B}
+  \frac{\alpha}{\sqrt{e^{ 8 \pi (M - D) \omega_{A} } + 1} } |1\rangle_{A}|1\rangle_{\bar A}|1\rangle_{B}|0\rangle_{\bar B}  \nonumber\\&&
\pm   \frac{\sqrt{1-\alpha^{2}}}{\sqrt{e^{- 8 \pi (M - D) \omega_{B} } + 1} } |1\rangle_{A}|0\rangle_{\bar A}|0\rangle_{B}|0\rangle_{\bar B}
\pm  \frac{\sqrt{1-\alpha^{2}}}{\sqrt{e^{ 8 \pi (M - D) \omega_{B} } + 1} } |1\rangle_{A}|0\rangle_{\bar A}|1\rangle_{B}|1\rangle_{\bar B}.
\end{eqnarray}
Here, the modes $\bar A$ and $\bar B$ are observed by hypothetical observers Anti-Alice and Anti-Bob inside the event horizon of the black hole, while $\omega_{A}$ and $\omega_{B}$ are the frequencies of modes $A$ and $B$, respectively.

Since the exterior region of the black hole is causally disconnected from its interior, Alice and Bob cannot detect physically inaccessible modes $\bar A$ and $\bar B$.  Consequently, we need to trace over these inaccessible modes to derive the density matrix for Alice and Bob
\begin{eqnarray}\label{w27}
\scalebox{0.76}{$
\rho^{{1},{\pm}}_{AB}=\left(\!\!\begin{array}{cccccccc}
 \frac{\alpha}{e^{- 8 \pi (M - D) \omega_{A} } + 1 } \frac{\alpha}{e^{- 8 \pi (M - D) \omega_{B} } + 1 }&0&0&\pm  \frac{\alpha}{\sqrt{e^{- 8 \pi (M - D) \omega_{A} } + 1} } \frac{\sqrt{1-\alpha^{2}}}{\sqrt{e^{- 8 \pi (M - D) \omega_{B} } + 1} }\\
0& \frac{\alpha}{e^{- 8 \pi (M - D) \omega_{A} } + 1 } \frac{\alpha}{e^{ 8 \pi (M - D) \omega_{B} } + 1 }&0&0\\
0&0& \frac{\alpha}{e^{ 8 \pi (M - D) \omega_{A} } + 1 } \frac{\alpha}{e^{- 8 \pi (M - D) \omega_{B} } + 1 }&0\\
\pm  \frac{\alpha}{\sqrt{e^{- 8 \pi (M - D) \omega_{A} } + 1} } \frac{\sqrt{1-\alpha^{2}}}{\sqrt{e^{- 8 \pi (M - D) \omega_{B} } + 1} }&0&0& \frac{\alpha}{e^{ 8 \pi (M - D) \omega_{A} } + 1 } \frac{\alpha}{e^{ 8 \pi (M - D) \omega_{B} } + 1 }+1-\alpha^{2}
\end{array}\!\!\right)
$},
\end{eqnarray}
\begin{eqnarray}\label{w28}
\scalebox{0.76}{$
\rho^{{2},{\pm}}_{AB}=\left(\!\!\begin{array}{cccccccc}
0&0&0&0\\
0& \frac{\alpha^{2}}{e^{- 8 \pi (M - D) \omega_{A} } + 1 } &\pm \frac{\alpha}{\sqrt{e^{- 8 \pi (M - D) \omega_{A} } + 1} } \frac{\sqrt{1-\alpha^{2}}}{\sqrt{e^{- 8 \pi (M - D) \omega_{B} } + 1} }&0\\
0&\pm \frac{\alpha}{\sqrt{e^{- 8 \pi (M - D) \omega_{A} } + 1} } \frac{\sqrt{1-\alpha^{2}}}{\sqrt{e^{- 8 \pi (M - D) \omega_{B} } + 1} }& \frac{1-\alpha^{2}}{e^{- 8 \pi (M - D) \omega_{B} } + 1 }&0\\
0&0&0& \frac{\alpha^{2}}{e^{ 8 \pi (M - D) \omega_{A} } + 1 }+\frac{1-\alpha^{2}}{e^{ 8 \pi (M - D) \omega_{B} } + 1 }
\end{array}\!\!\right)
$}.
\end{eqnarray}

In this paper, we utilize the concurrence as a metric to quantify quantum entanglement in the dilaton black hole. For a mixed two-qubit state, its concurrence can be defined as
\begin{eqnarray}\label{w29}
 C(\rho)=\max\{0,\lambda_{1}-\lambda_{2}-\lambda_{3}-\lambda_{4}\}, \quad \lambda_{i} \geq \lambda_{i+1} \geq 0 ,
\end{eqnarray}
where $\lambda_{i}$ are the square roots of the eigenvalues of the matrix $\rho\widetilde{\rho}$ with the ``spin-flip" matrix $\widetilde{\rho} = (\sigma_{y}\otimes\sigma_{y})\rho^{\ast}(\sigma_{y}\otimes\sigma_{y})$ \cite{Q1}.
Employing Eq.(\ref{w29}), we obtain the analytical expressions of the concurrence for Eqs.(\ref{w27}) and (\ref{w28}) as
\begin{eqnarray}\label{w30}
C(\rho^{{1},{\pm}}_{AB})&=&\max\bigg\{0, \frac{2\alpha}{\sqrt{e^{- 8 \pi (M - D) \omega_{A} } + 1} } \frac{\alpha}{\sqrt{e^{- 8 \pi (M - D) \omega_{B} } + 1} }  \bigg( \frac{\sqrt{1-\alpha^{2}}}{\alpha}
\nonumber\\&&
- \frac{1}{\sqrt{e^{ 8 \pi (M - D) \omega_{A} } + 1} } \frac{1}{\sqrt{e^{8 \pi (M - D) \omega_{B} } + 1} } \bigg) \bigg\},
\end{eqnarray}
\begin{eqnarray}\label{w31}
C(\rho^{{2},{\pm}}_{AB})=\max\{0, \frac{2\alpha}{\sqrt{e^{- 8 \pi (M - D) \omega_{A} } + 1} } \frac{\sqrt{1-\alpha^{2}}}{\sqrt{e^{- 8 \pi (M - D) \omega_{B} } + 1} } \}.
\end{eqnarray}

In order to compare quantum entanglement and coherence, we introduce two measures of quantum coherence: the $l_{1}$-norm of coherence and the relative entropy of coherence (REC) \cite{L47}. 
In a reference basis $\{|i⟩\}_{i=1,...,n}$ of a $n$-dimensional system, the $l_{1}$-norm of quantum coherence can be defined as the sum of the absolute values of all the off-diagonal elements of the system density
matrix $\rho$,
\begin{eqnarray}\label{q1}
C_{\mathrm{l_{1}}}(\rho)=\sum_{i \neq j}|\rho_{i,j}|.
\end{eqnarray}
Employing Eq.(\ref{q1}), we obtain the analytical expressions of the $l_{1}$-norm of quantum coherence for Eqs.(\ref{w27}) and (\ref{w28}) as
\begin{eqnarray}\label{q2}
C_{\mathrm{l_{1}}}(\rho^{{1},{\pm}}_{AB})=C_{\mathrm{l_{1}}}(\rho^{{2},{\pm}}_{AB})= \frac{2\alpha}{\sqrt{e^{- 8 \pi (M - D) \omega_{A} } + 1} } \frac{\sqrt{1-\alpha^{2}}}{\sqrt{e^{- 8 \pi (M - D) \omega_{B} } + 1} }.
\end{eqnarray}

The measure of the REC can be written as
\begin{eqnarray}\label{q3}
C_{\mathrm{REC}}(\rho)=S(\rho_{diag})-S(\rho),
\end{eqnarray}
where $S(\rho)$ indicates the von Neumann entropy of quantum state $\rho$, and $\rho_{diag}$ is the state obtained from $\rho$ by removing all off-diagonal elements. 
For a mixed two-qubit state $\rho_{AB}$, its REC can be defined as
\begin{eqnarray}\label{q4}
C_{\mathrm{REC}}(\rho_{AB})=-\sum_{i}\gamma_{i} \mathrm{log}_{2} \gamma_{i} + \sum_{j}\lambda_{j} \mathrm{log}_{2} \lambda_{j},
\end{eqnarray}
where $\gamma_{i}$ are the diagonal elements of $\rho_{AB}$ and $\lambda_{j}$ are nonzero eigenvalues of quantum state $\rho_{AB}$.
For this purpose, we need to calculate the eigenvalues of the density matrices $\rho^{{1},{\pm}}_{AB}$ and $\rho^{{2},{\pm}}_{AB}$.
The density matrix of $\rho^{{1},{\pm}}_{AB}$ has four nonzero eigenvalues
\begin{align*}
\lambda_1^{1, \pm} &= \rho_{22}^{1, \pm} =  \frac{\alpha}{e^{- 8 \pi (M - D) \omega_{A} } + 1 } \frac{\alpha}{e^{ 8 \pi (M - D) \omega_{B} } + 1 },\\
\lambda_2^{1, \pm} &= \rho_{33}^{1, \pm} =  \frac{\alpha}{e^{- 8 \pi (M - D) \omega_{B} } + 1 } \frac{\alpha}{e^{ 8 \pi (M - D) \omega_{A} } + 1 },\\
\lambda_3^{1, \pm} &= \frac{1}{2} \bigg( \rho_{11}^{1, \pm} + \rho_{44}^{1, \pm} - \sqrt{(\rho_{11}^{1, \pm})^2 + 4(\rho_{14}^{1, \pm})^2 - 2\rho_{11}^{1, \pm} \rho_{44}^{1, \pm} + (\rho_{44}^{1, \pm})^2} \bigg) \\
&= \begin{aligned}[t]
   &\frac{1}{2} \bigg\{ 1 - \alpha^2 +  \frac{\alpha}{e^{- 8 \pi (M - D) \omega_{A} } + 1 } \frac{\alpha}{e^{- 8 \pi (M - D) \omega_{B} } + 1 } +  \frac{\alpha}{e^{ 8 \pi (M - D) \omega_{A} } + 1 } \frac{\alpha}{e^{ 8 \pi (M - D) \omega_{B} } + 1 } \\
   &\quad - \bigg[  \frac{\alpha^2}{(e^{- 8 \pi (M - D) \omega_{A} } + 1)^2 } \frac{\alpha^2}{(e^{- 8 \pi (M - D) \omega_{B} } + 1)^2 } + \bigg (1 - \alpha^2 +  \frac{\alpha}{e^{ 8 \pi (M - D) \omega_{A} } + 1 } \frac{\alpha}{e^{ 8 \pi (M - D) \omega_{B} } + 1 } \bigg)^2 \\
   &\quad -  \frac{2\alpha}{e^{- 8 \pi (M - D) \omega_{A} } + 1 } \frac{\alpha}{e^{- 8 \pi (M - D) \omega_{B} } + 1 } \bigg (-1 + \alpha^2 +  \frac{\alpha}{e^{ 8 \pi (M - D) \omega_{A} } + 1 } \frac{\alpha}{e^{ 8 \pi (M - D) \omega_{B} } + 1 }\bigg ) \bigg]^{\frac{1}{2}} \bigg\},
   \end{aligned} \\
\lambda_4^{1, \pm} &= \frac{1}{2} \bigg( \rho_{11}^{1, \pm} + \rho_{44}^{1, \pm} + \sqrt{(\rho_{11}^{1, \pm})^2 + 4(\rho_{14}^{1, \pm})^2 - 2\rho_{11}^{1, \pm} \rho_{44}^{1, \pm} + (\rho_{44}^{1, \pm})^2} \bigg) \\
&= \begin{aligned}[t]
   &\frac{1}{2} \bigg\{ 1 - \alpha^2 +  \frac{\alpha}{e^{- 8 \pi (M - D) \omega_{A} } + 1 } \frac{\alpha}{e^{- 8 \pi (M - D) \omega_{B} } + 1 }  +  \frac{\alpha}{e^{ 8 \pi (M - D) \omega_{A} } + 1 } \frac{\alpha}{e^{ 8 \pi (M - D) \omega_{B} } + 1 } \\
   &\quad + \bigg[  \frac{\alpha^2}{(e^{- 8 \pi (M - D) \omega_{A} } + 1)^2 } \frac{\alpha^2}{(e^{- 8 \pi (M - D) \omega_{B} } + 1)^2 } 
    + \bigg(1 - \alpha^2 +  \frac{\alpha}{e^{ 8 \pi (M - D) \omega_{A} } + 1 } \frac{\alpha}{e^{ 8 \pi (M - D) \omega_{B} } + 1 }\bigg)^2 \\
   &\quad -  \frac{2\alpha}{e^{- 8 \pi (M - D) \omega_{A} } + 1 } \frac{\alpha}{e^{- 8 \pi (M - D) \omega_{B} } + 1 } \bigg(-1 + \alpha^2 
    +  \frac{\alpha}{e^{ 8 \pi (M - D) \omega_{A} } + 1 } \frac{\alpha}{e^{ 8 \pi (M - D) \omega_{B} } + 1 }\bigg) \bigg]^{\frac{1}{2}} \bigg\},
   \end{aligned}
\end{align*}
where $\rho_{11}^{1, \pm}$, $\rho_{14}^{1, \pm}$, $\rho_{22}^{1, \pm}$, $\rho_{33}^{1, \pm}$ and $\rho_{44}^{1, \pm}$ are the matrix elements of $\rho^{{1},{\pm}}_{AB}$. 
In terms of the density matrix of $\rho^{{2},{\pm}}_{AB}$, it has two nonzero eigenvalues
\begin{align*}
\lambda_1^{2, \pm} &= \frac{1}{2} \bigg(\rho_{22}^{{2},{\pm}}+\rho_{33}^{{2},{\pm}}+\sqrt{(\rho_{22}^{{2},{\pm}})^2+4(\rho_{23}^{{2},{\pm}})^2-2\rho_{22}^{{2},{\pm}}\rho_{33}^{{2},{\pm}}+(\rho_{33}^{{2},{\pm}})^2} \bigg) \\
&= \begin{aligned}[t]
   & \frac{\alpha^{2}}{e^{ 8 \pi (M - D) \omega_{A} } + 1 } +  \frac{1-\alpha^{2}}{e^{ 8 \pi (M - D) \omega_{B} } + 1 },
   \end{aligned} \\
\lambda_2^{2, \pm} &= \rho^{{2},{\pm}}_{44}= \frac{\alpha^{2}}{e^{- 8 \pi (M - D) \omega_{A} } + 1 } +  \frac{1-\alpha^{2}}{e^{- 8 \pi (M - D) \omega_{B} } + 1 },
\end{align*}
where $\rho_{22}^{2, \pm}$, $\rho_{23}^{2, \pm}$, $\rho_{33}^{2, \pm}$ and $\rho_{44}^{2, \pm}$ are the matrix elements of $\rho^{{2},{\pm}}_{AB}$.

\begin{figure}
\begin{minipage}[t]{0.5\linewidth}
\centering
\includegraphics[width=3.0in,height=5.2cm]{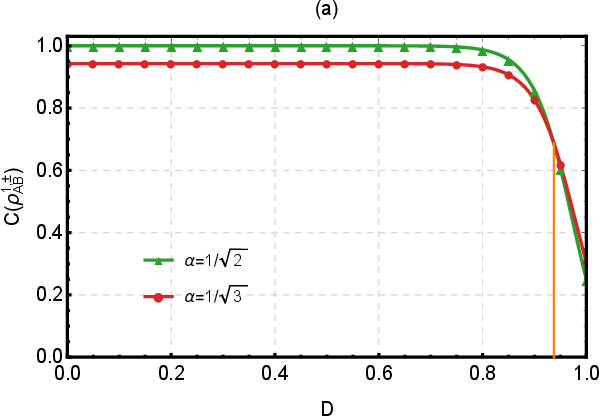}
\label{fig1a}
\end{minipage}%
\begin{minipage}[t]{0.5\linewidth}
\centering
\includegraphics[width=3.0in,height=5.2cm]{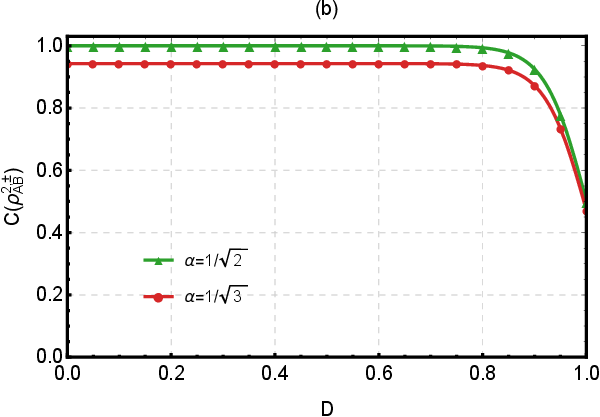}
\label{fig1b}
\end{minipage}%

\begin{minipage}[t]{0.5\linewidth}
\centering
\includegraphics[width=3.0in,height=5.2cm]{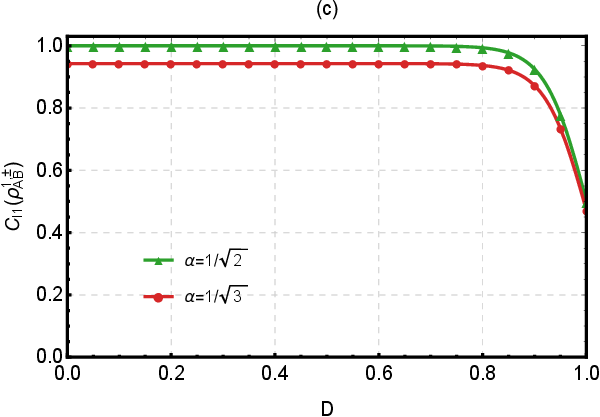}
\label{fig1c}
\end{minipage}%
\begin{minipage}[t]{0.5\linewidth}
\centering
\includegraphics[width=3.0in,height=5.2cm]{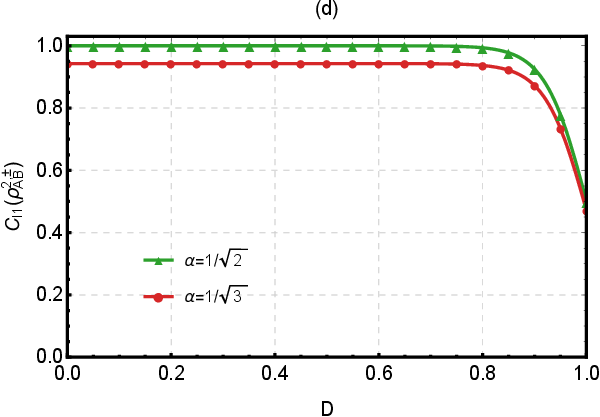}
\label{fig1d}
\end{minipage}%

\begin{minipage}[t]{0.5\linewidth}
\centering
\includegraphics[width=3.0in,height=5.2cm]{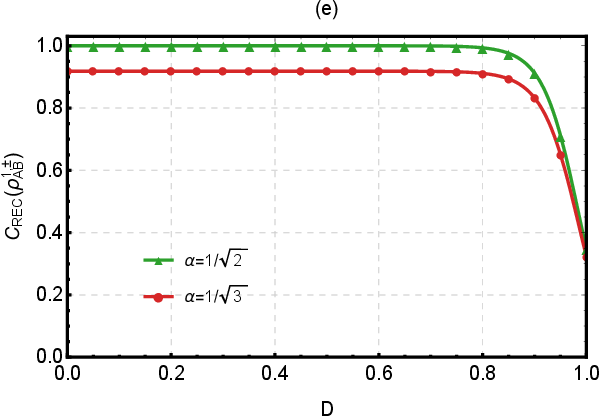}
\label{fig1e}
\end{minipage}%
\begin{minipage}[t]{0.5\linewidth}
\centering
\includegraphics[width=3.0in,height=5.2cm]{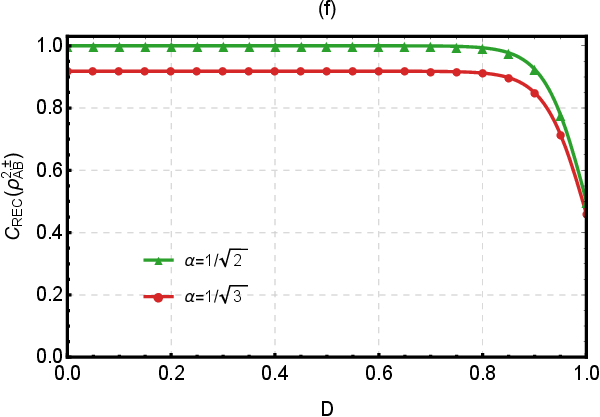}
\label{fig1f}
\end{minipage}%
\caption{The concurrence $C(\rho^{{1},{\pm}}_{AB})$ and $C(\rho^{{2},{\pm}}_{AB})$, the $l_{1}$-norm of coherence $C_{\mathrm{l_{1}}}(\rho^{{1},{\pm}}_{AB})$ and $C_{\mathrm{l_{1}}}(\rho^{{2},{\pm}}_{AB})$, the REC $C_{\mathrm{REC}}(\rho^{{1},{\pm}}_{AB})$ and $C_{\mathrm{REC}}(\rho^{{2},{\pm}}_{AB})$ as functions of the dilaton $D$ for different initial parameters $\alpha$ with fixed $M = \omega_{A} = \omega_{B} = 1$.}
\label{Fig1}
\end{figure}

Fig.\ref{Fig1} shows the evolution of the concurrence, the $l_{1}$-norm of coherence, and the REC  for the four classes of Bell-like states $|\phi^{{1},{\pm}}_{AB}\rangle$ and $|\Psi^{{2},{\pm}}_{AB}\rangle$, as functions of the dilaton $D$ (with $M = \omega_{A} = \omega_{B} = 1$). 
For both the maximally entangled ($\alpha=\frac{1}{\sqrt{2}}$) and non-maximally entangled ($\alpha=\frac{1}{\sqrt{3}}$) initial states,  all three quantities exhibit a monotonic decrease with the increase of the dilaton $D$, directly manifesting the degrading influence of the gravitational effect on quantum resources. 
Fig.\ref{Fig1}(a) reveals a notable crossover for the $|\phi^{{1},{\pm}}_{AB}\rangle$ states. While the concurrence of the maximally entangled state is initially higher, it decays more rapidly with increasing $D$. Beyond a critical dilaton value $D = 1-\frac{1}{8\pi}\ln(2\sqrt{2}+2)$, the non-maximally entangled state retains stronger entanglement. 
This counterintuitive result demonstrates that under strong gravitational effects, the non-maximally entangled states can outperform maximally entangled states, challenging the previous conclusions in the dilaton black hole that the maximally entangled states always guarantee superior robustness \cite{J44,J45,J46,J47,J48,J49,J50,J51,J52,J53,J54,J55,J56,J57,J58,J59,J60}.
In contrast, for the $|\Psi^{{2},{\pm}}_{AB}\rangle$ states shown in Fig.\ref{Fig1}(b), the maximally entangled state maintains a higher concurrence throughout the degradation, consistent with the analytic expression in Eq.(\ref{w31}).
Furthermore, comparison between Figs.\ref{Fig1}(a) and (b) indicates that the $|\Psi^{{2},{\pm}}_{AB}\rangle$ states generally exhibits greater entanglement resilience against the Hawking effect than the $|\phi^{{1},{\pm}}_{AB}\rangle$ states.

The $l_{1}$-norm of coherence, given by Eq.(\ref{q2}), is identical for both state families at a given $\alpha$ and decreases monotonically as the dilaton $D$ increases. 
Notably, the state with larger initial coherence ($\alpha=\frac{1}{\sqrt{2}}$) consistently maintains stronger coherence under the influence of the dilaton black hole in Figs.\ref{Fig1}(c) and (d).
This monotonic relationship underscores that quantum coherence degrades primarily governed by its initial magnitude, making it a more uniformly stable quantum resource compared to entanglement in gravitational backgrounds. Although the $l_{1}$-norm of coherence fails to distinguish the Bell-like families at the same $\alpha$, their $C_{\mathrm{REC}}$ shows significant differences in Figs.\ref{Fig1}(e) and (f). 
This indicates that different measures of quantum coherence capture distinct structural features of quantum states and respond differently to gravitational effects, which ultimately stems from the inherent inconsistency in the definitions and physical meanings of these coherence measures. Our analysis reveals that no single class of initial states is universally optimal in curved spacetime. Instead, the suitability of a given initial quantum state near a dilaton black hole is strongly resource-dependent, varying according to whether quantum entanglement or quantum coherence is the relevant operational quantity. This resource-specific behavior highlights the nontrivial interplay between gravity and distinct quantum correlations, and underscores the necessity of tailoring state preparation strategies to the targeted quantum information task in gravitational environments.

\begin{figure}
\begin{minipage}[t]{0.5\linewidth}
\centering
\includegraphics[width=3.0in,height=5.2cm]{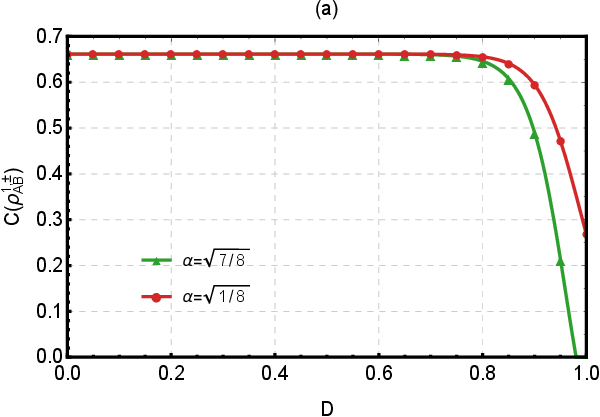}
\label{fig2a}
\end{minipage}%
\begin{minipage}[t]{0.5\linewidth}
\centering
\includegraphics[width=3.0in,height=5.2cm]{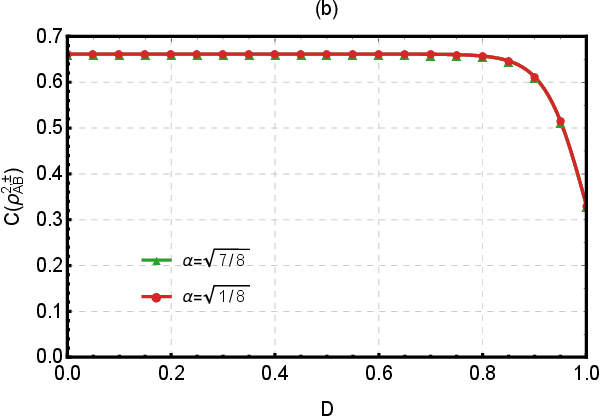}
\label{fig2b}
\end{minipage}%

\begin{minipage}[t]{0.5\linewidth}
\centering
\includegraphics[width=3.0in,height=5.2cm]{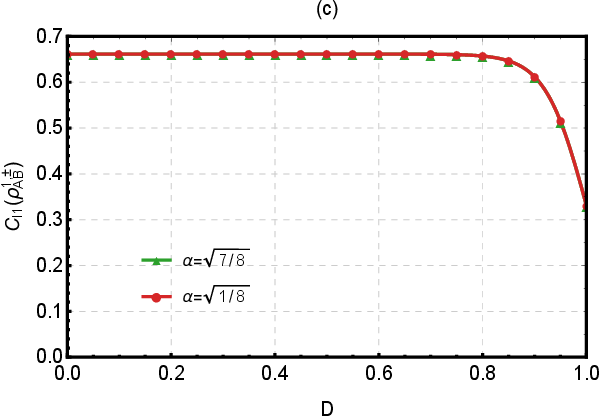}
\label{fig2c}
\end{minipage}%
\begin{minipage}[t]{0.5\linewidth}
\centering
\includegraphics[width=3.0in,height=5.2cm]{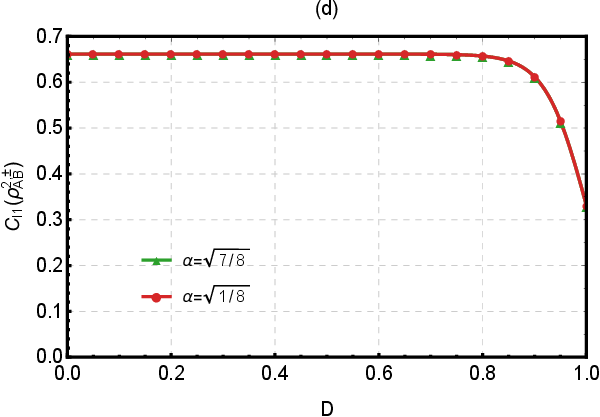}
\label{fig2d}
\end{minipage}%

\begin{minipage}[t]{0.5\linewidth}
\centering
\includegraphics[width=3.0in,height=5.2cm]{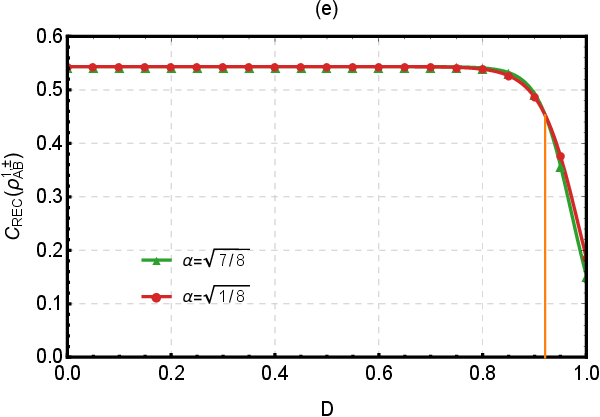}
\label{fig2e}
\end{minipage}%
\begin{minipage}[t]{0.5\linewidth}
\centering
\includegraphics[width=3.0in,height=5.2cm]{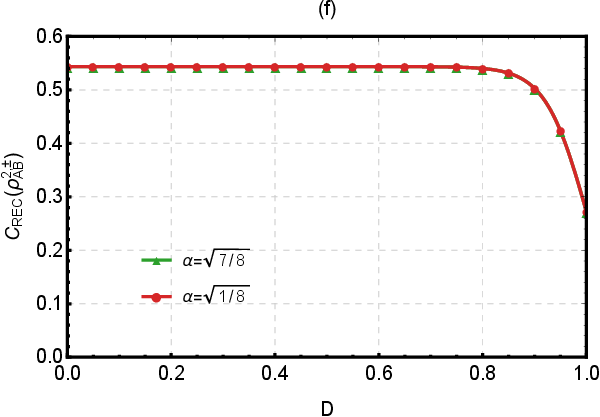}
\label{fig2f}
\end{minipage}%
\caption{The concurrence $C(\rho^{{1},{\pm}}_{AB})$ and $C(\rho^{{2},{\pm}}_{AB})$, the $l_{1}$-norm of coherence $C_{\mathrm{l_{1}}}(\rho^{{1},{\pm}}_{AB})$ and $C_{\mathrm{l_{1}}}(\rho^{{2},{\pm}}_{AB})$, the REC $C_{\mathrm{REC}}(\rho^{{1},{\pm}}_{AB})$ and $C_{\mathrm{REC}}(\rho^{{2},{\pm}}_{AB})$ as functions of the dilaton $D$ for different initial parameters $\alpha$ with fixed $M = \omega_{A} = \omega_{B} = 1$.}
\label{Fig2}
\end{figure}

Fig.\ref{Fig2} further illustrates the effects of Hawking radiation on both entanglement and coherence by comparing two states with parameters $\alpha=1/\sqrt{8}$ and $\alpha=\sqrt{7/8}$, which are prepared to have identical initial concurrence and coherence in  dilaton spacetime.
For the $|\phi^{{1},{\pm}}_{AB}\rangle$ family, the concurrence for $\alpha=\sqrt{\frac{7}{8}}$ undergoes sudden death as the dilaton $D$ increases, whereas for $\alpha=\frac{1}{\sqrt{8}}$ it survives indefinitely in the context of extreme black holes. 
This dramatic divergence directly validates the condition from Eq.(\ref{w30}): the concurrence $C(\rho^{{1},{\pm}}_{AB})$ survival depends critically on whether $\frac{\sqrt{1-\alpha^2}}{\alpha}>\frac{1}{\sqrt{e^{ 8 \pi (M - D) \omega_{A} } + 1} } \frac{1}{\sqrt{e^{ 8 \pi (M - D) \omega_{B} } + 1} }$.
Conversely, for the $|\Psi^{{2},{\pm}}_{AB}\rangle$ states, the concurrence can persist indefinitely and is identical for the two chosen $\alpha$ values, as per Eq.(\ref{w31}).
Therefore, this reveals a clear classification of fermionic Bell-like states based on their entanglement resilience in curved spacetime.
Regarding quantum coherence, both $C_{\mathrm{l_{1}}}$ and $C_{\mathrm{REC}}$ again exhibit a monotonic decay with increasing $D$.  This implies that, as long as a quantum state carries nonzero initial coherence in flat spacetime, such coherence can persist indefinitely in a dilaton spacetime background without undergoing complete extinction. 
Notably, while the \( l_1 \)-norm of coherence remains identical for the four Bell-like states at a given $\alpha$, their $C_{\mathrm{REC}}$ shows significant differences.
In summary, under the exchange of the parameters $\alpha$ and $\sqrt{1-\alpha^{2}}$
(i.e., $\alpha \leftrightarrow \sqrt{1-\alpha^{2}}$), the Hawking effect of the black hole
breaks the symmetry of quantum entanglement and REC,
whereas the symmetry of the $l_{1}$-norm of coherence remains completely preserved. Therefore, by exploiting different types of quantum resources, we gain a more comprehensive and refined understanding of how dilaton black holes can be probed through quantum effects.

\section{Conclusions}
In this work, we have analyzed how the Hawking radiation of the GHS dilaton black hole influences quantum entanglement and quantum coherence for four distinct classes of Bell-like fermionic states. The scenario involves two observers, Alice and Bob, who share an initial Bell-like state in flat spacetime and subsequently remain stationary outside the event horizon of the dilaton black hole.  Our findings reveal
several important physical insights: \textbf{(i) non-maximally entangled states can outperform maximally entangled ones:} we find that, under the influence of the dilaton-induced Hawking effect, certain non-maximally entangled states retain higher entanglement than maximally entangled states. This overturns the common intuition that maximal quantum resources always offer the greatest robustness in dilaton spacetime \cite{J44,J45,J46,J47,J48,J49,J50,J51,J52,J53,J54,J55,J56,J57,J58,J59,J60};  \textbf{(ii) quantum coherence exhibits a distinct  behavior:} unlike entanglement, quantum coherence shows a monotonic advantage: states with larger initial coherence consistently maintain stronger coherence during their evolution outside the dilaton black hole. This demonstrates that coherence behaves as a more uniformly stable resource compared to entanglement in gravitational backgrounds; \textbf{(iii) Hawking radiation induces qualitatively different entanglement decay patterns:} two of the Bell-like states undergo entanglement sudden death as the dilaton parameter increases, while the remaining two families preserve a finite amount of entanglement for all parameter values. This reveals a clear classification of fermionic Bell-like states based on their entanglement resilience in curved spacetime; \textbf{(iv) different coherence measures distinguish the four Bell-like families:} although the four classes of Bell-like states share identical coherence when quantified by the \( l_1 \)-norm, their REC shows significant differences. This highlights that various coherence measures probe different structural aspects of the states and respond differently to gravitational effects.
In summary, our analysis reveals that gravitational effects act highly unevenly on different quantum resources and on distinct classes of Bell-like states. These findings highlight the necessity of choosing the most appropriate initial states for a given quantum resource when designing quantum information protocols in the background of a dilaton black hole.

\begin{acknowledgments}
This work is supported by the Joint Funds Program of Liaoning Provincial Natural Science Foundation (Grant No. 2025-BSLH-223) and Specific Fund of Fundamental Scientific Research Operating Expenses for Undergraduate Universities in Liaoning Province under (Grant No. LJ212510165012).	
\end{acknowledgments}


\end{document}